\documentclass[aps,prd,showpacs,amssymb,superscriptaddress,nofootinbib,twocolumn]{revtex4-1}

\usepackage{amsmath,amsfonts,amsthm,amssymb}
\usepackage{braket}
\usepackage{setspace}
\usepackage{amsmath}
\usepackage{appendix}
\usepackage[ansinew]{inputenc}
\usepackage{bbm}
\usepackage{bm}
\usepackage{amsbsy}
\usepackage{dsfont} 
\usepackage{graphicx} 
\usepackage{epsfig}
\usepackage{epstopdf}
\usepackage{dsfont}
\usepackage{color}
\usepackage[colorlinks]{hyperref}
\usepackage[figure,table]{hypcap}
\usepackage{enumerate}
\usepackage[resetlabels]{multibib}

\hypersetup{
	bookmarksnumbered,
	pdfstartview={FitH},
	citecolor={darkgreen},
	linkcolor={darkred},
	urlcolor={darkblue},
	pdfpagemode={UseOutlines}}
\definecolor{darkgreen}{RGB}{50,190,50}
\definecolor{darkblue}{RGB}{0,0,190}
\definecolor{darkred}{RGB}{238,0,0}
\usepackage{soul}

\newcommand{\USB}{\ensuremath{U_{\hspace{-1.3pt}\protect\raisebox{0pt}{\tiny{$S\hspace*{-0.7pt}B$}}}}}
\newcommand{\SB}{\ensuremath{_{\hspace{-0.5pt}\protect\raisebox{0pt}{\tiny{$S\hspace*{-0.5pt}B$}}}}}
\newcommand{\Sys}{\ensuremath{_{\hspace{-0.5pt}\protect\raisebox{0pt}{\tiny{$S$}}}}}
\newcommand{\HI}{\ensuremath{H_{\hspace{-0.5pt}\protect\raisebox{0pt}{\tiny{$I$}}}}}
\newcommand{\bath}{\ensuremath{_{\hspace{-0.5pt}\protect\raisebox{0pt}{\tiny{$B$}}}}}
\newcommand{\Sone}{\ensuremath{_{\hspace{-0.5pt}\protect\raisebox{0pt}{\tiny{$S_{1}$}}}}}
\newcommand{\Stwo}{\ensuremath{_{\hspace{-0.5pt}\protect\raisebox{0pt}{\tiny{$S_{2}$}}}}}
\newcommand{\Si}{\ensuremath{_{\hspace{-0.5pt}\protect\raisebox{0pt}{\tiny{$S_{i}$}}}}}
\newcommand{\Sonetwo}{\ensuremath{_{\hspace{-0.5pt}\protect\raisebox{0pt}{\tiny{$S_{1}S_{2}$}}}}}
\newcommand{\Soneortwo}{\ensuremath{_{\hspace{-0.5pt}\protect\raisebox{0pt}{\tiny{$S_{1}(S_{2})$}}}}}

\newcommand{\Stwoorone}{\ensuremath{_{\hspace{-0.5pt}\protect\raisebox{0pt}{\tiny{$S_{2}(S_{1})$}}}}}

\newcommand{\dvline}{\ensuremath{\hspace*{0.5pt}|\hspace*{-1pt}|\hspace*{0.5pt}}}
\newcommand{\sigone}{\ensuremath{\sigma^{\hspace{-0.0pt}\protect\raisebox{0pt}{\tiny{$S_{1}$}}}}}
\newcommand{\sigtwo}{\ensuremath{\sigma^{\hspace{-0.0pt}\protect\raisebox{0pt}{\tiny{$S_{2}$}}}}}
\newcommand{\sigi}{\ensuremath{\sigma^{\hspace{-0.0pt}\protect\raisebox{0pt}{\tiny{$S_{i}$}}}}}

\newcommand{\rhoI}{\ensuremath{\rho_{\protect\raisebox{-0.5pt}{\hspace*{0.5pt}\tiny{$\mathrm{I}$}}}}}
\newcommand{\suprhoI}{\ensuremath{^{(\rho_{\protect\raisebox{-1pt}{\hspace*{0.5pt}\tiny{$\mathrm{I}$}}}\hspace*{-0.5pt})}}}
\newcommand{\alphaI}{\ensuremath{\alpha_{\protect\raisebox{-0.5pt}{\hspace*{0.5pt}\tiny{$\mathrm{I}$}}}}}
\newcommand{\rhoII}{\ensuremath{\rho_{\protect\raisebox{-0.5pt}{\hspace*{0.5pt}\tiny{$\mathrm{I\hspace*{-0.5pt}I}$}}}}}
\newcommand{\suprhoII}{\ensuremath{^{(\rho_{\protect\raisebox{-1pt}{\hspace*{0.5pt}\tiny{$\mathrm{I\hspace*{-0.5pt}I}$}}}\hspace*{-0.5pt})}}}
\newcommand{\alphaII}{\ensuremath{\alpha_{\protect\raisebox{-0.5pt}{\hspace*{0.5pt}\tiny{$\mathrm{I\hspace*{-0.5pt}I}$}}}}}

\newcommand{\aone}{\ensuremath{a_{\protect\raisebox{-1.5pt}{\tiny{1}}}}}
\newcommand{\aonedg}{\ensuremath{a_{\protect\raisebox{-0.5pt}{\tiny{1}}}^{\dagger}}}
\newcommand{\atwo}{\ensuremath{a_{\protect\raisebox{-1.5pt}{\tiny{2}}}}}
\newcommand{\atwodg}{\ensuremath{a_{\protect\raisebox{-0.5pt}{\tiny{2}}}^{\dagger}}}
\newcommand{\an}[1]{\ensuremath{a_{\protect\raisebox{-1.5pt}{\scriptsize{$#1$}}}}}
\newcommand{\andg}[1]{\ensuremath{a_{\protect\raisebox{-0.5pt}{\scriptsize{$#1$}}}^{\dagger}}}

\newcommand{\bone}{\ensuremath{b_{\protect\raisebox{-1.5pt}{\tiny{1}}}}}
\newcommand{\bonedg}{\ensuremath{b_{\protect\raisebox{-0.5pt}{\tiny{1}}}^{\dagger}}}
\newcommand{\btwo}{\ensuremath{b_{\protect\raisebox{-1.5pt}{\tiny{2}}}}}
\newcommand{\btwodg}{\ensuremath{b_{\protect\raisebox{-0.5pt}{\tiny{2}}}^{\dagger}}}
\newcommand{\bn}[1]{\ensuremath{b_{\protect\raisebox{-1.5pt}{\scriptsize{$#1$}}}}}
\newcommand{\bndg}[1]{\ensuremath{b_{\protect\raisebox{-0.5pt}{\scriptsize{$#1$}}}^{\dagger}}}

\newcommand{\cone}{\ensuremath{c_{\protect\raisebox{-1.5pt}{\tiny{1}}}}}
\newcommand{\conedg}{\ensuremath{c_{\protect\raisebox{-0.5pt}{\tiny{1}}}^{\dagger}}}
\newcommand{\ctwo}{\ensuremath{c_{\protect\raisebox{-1.5pt}{\tiny{2}}}}}
\newcommand{\ctwodg}{\ensuremath{c_{\protect\raisebox{-0.5pt}{\tiny{2}}}^{\dagger}}}
\newcommand{\cn}[1]{\ensuremath{c_{\protect\raisebox{-1.5pt}{\scriptsize{$#1$}}}}}
\newcommand{\cndg}[1]{\ensuremath{c_{\protect\raisebox{-0.5pt}{\scriptsize{$#1$}}}^{\dagger}}}

\newcommand{\fbra}[1]{\ensuremath{\left\langle\!\left\langle\right.\right.\! #1 \!\left.\left.\right|\hspace*{-0.75pt}\right|}}
\newcommand{\fket}[1]{\ensuremath{\left|\hspace*{-0.75pt}\left|\right.\right.\! #1 \!\left.\left.\right\rangle\!\right\rangle}}

\newcommand{\comm}[2]{\ensuremath{\left[\right.\! #1 \,, #2 \!\left.\right]}}

\newcommand{\idN}[1]{\ensuremath{\mathds{1}_{\hspace*{-1.0pt}\protect\raisebox{-1.0pt}{\scriptsize{$ #1 $}}}}}

\newcommand{\subA}[1]{\ensuremath{_{\hspace*{-1.0pt}\protect\raisebox{-1.0pt}{\tiny{$ #1 $}}}}}

\DeclareMathOperator{\artanh}{artanh}

\DeclareMathOperator{\diag}{diag}

\newcommand{\tr}{\textnormal{Tr}}
\newcommand{\djj}{d\kern-0.4em\char"16\kern-0.1em}

\begin{document}

\title{Energetics of correlations in interacting systems}
\author{Nicolai Friis}
\email{nicolai.friis@uibk.ac.at}
\affiliation{
Institute for Theoretical Physics, University of Innsbruck,
Technikerstra{\ss}e 21a,
A-6020 Innsbruck,
Austria}
\author{Marcus Huber}
\email{marcus.huber@univie.ac.at}
\affiliation{Group of Applied Physics, University of Geneva, 1211 Geneva 4, Switzerland}
\affiliation{Departament de F\'isica, Universitat Aut\`onoma de Barcelona, 08193 Bellaterra, Spain}
\affiliation{ICFO - Institut de Ciencies Fotoniques, The Barcelona Institute of Science and Technology, 08860 Castelldefels (Barcelona), Spain}
\author{Mart{\'i} Perarnau-Llobet}
\email{marti.perarnau@icfo.es}
\affiliation{ICFO - Institut de Ciencies Fotoniques, The Barcelona Institute of Science and Technology, 08860 Castelldefels (Barcelona), Spain}

\begin{abstract}
A fundamental connection between thermodynamics and information theory arises from the fact that correlations exhibit an inherent work value. For noninteracting systems this translates to a work cost for establishing correlations. Here we investigate the relationship between work and correlations in the presence of interactions that cannot be controlled or removed. For such naturally coupled systems, which are correlated even in thermal equilibrium, we determine general strategies that can reduce the work cost of correlations, and illustrate these for a selection of exemplary physical systems.
\end{abstract}

\maketitle

\section{Introduction}

Quantum information (QI) and quantum thermodynamics (QT) can both be framed as resource theories~\cite{CoeckeFritzSpekkens2016}. Based on the fundamental laws of quantum physics, these theories describe the (minimal) resources needed to perform certain tasks of interest. In order to identify the relevant resources, one first determines which states and operations are freely available, taking into account practical limitations on physical operations. Within QT, thermal systems and energy preserving operations are considered to be ``for free", whereas systems out of equilibrium and operations that require external energy constitute resources~\cite{BrandaoHorodeckiOppenheimRenesSpekkens2013,NavascuesGarciaPintos2015}. In QI, on the other hand, the paradigmatic task is efficient communication. In this context one assumes local operations and classical communication (LOCC) to be for free, whereas entangled quantum systems are resources that enable tasks beyond the restrictions of LOCC~\cite{HorodeckiOppenheim2013a,EltschkaSiewert2014}.

Both of these resource theories can be considered to be simplifications of a more general physical framework: In either case only the restrictions of one area are taken into account. However, especially in quantum systems limitations from both thermodynamics and information theory present themselves simultaneously, which has greatly stimulated investigations of the connection between QI and QT (see Refs.~\cite{GooldHuberRieraDelRioSkrzypczyk2016,MillenXuereb2016,VinjanampathyAnders2015} for recent reviews). For instance, locality restrictions on the allowed operations can limit the efficiency of thermodynamic processes~\cite{OppenheimHorodeckiMPR2002,Zurek2003, AlickiFannes2013,HovhannisyanPerarnauLlobetHuberAcin2013,WilmingGallegoEisert2016,BinderVinjanampathyModiGoold2015}, while correlations can enhance the performance of thermodynamic tasks~\cite{Zurek2003,DelRioAbergRennerDahlstenVedral2011,SagawaUeda2008,FunoWatanabeUeda2013,BrunnerHuberLindenPopescuSilvaSkrzypczyk2014,
PerarnauLlobetHovhannisyanHuberSkrzypczykBrunnerAcin2015,BragaRulliOliveiraSarandy2014,LostaglioMuellerPastena2015}, and may even change the natural direction of the heat flow~\cite{JenningsRudolph2010,Partovi2008}. Conversely, a nonzero ambient temperature induces a nonzero entropy, which limits the capacity for establishing (quantum) correlations~\cite{HuberPerarnauHovhannisyanSkrzypczykKloecklBrunnerAcin2015,BruschiPerarnauLlobetFriisHovhannisyanHuber2015}. To overcome the constraints of LOCC, QI tasks hence require a supply of thermodynamic resources in the form of free energy.

Here, we aim to study the exchange between energy and correlations in the particularly transparent setting considered in Refs.~\cite{HuberPerarnauHovhannisyanSkrzypczykKloecklBrunnerAcin2015,BruschiPerarnauLlobetFriisHovhannisyanHuber2015}: Given a collection of uncorrelated thermal states at the same temperature~$T$, one is interested in determining the minimal energy~$W$ that is needed to create (quantum) correlations. For a bipartite system with access to an auxiliary thermal bath at temperature~$T$, one finds the relation~\cite{SagawaUeda2008,JevticJenningsRudolph2012a,BruschiPerarnauLlobetFriisHovhannisyanHuber2015,EspositoVanDenBroeck2011}
\begin{align}
    W   &\geq\,T\Delta\mathcal{I}\Sonetwo\,,
    \label{WTI}
\end{align}
where $\Delta\mathcal{I}\Sonetwo$ is the gain of correlations between the subsystems $S_{1}$ and $S_{2}$ as measured by the mutual information, and we note that we work in units such that $\hbar=k_{\protect\raisebox{-0pt}{\tiny{B}}}=1$ throughout this paper. The expression in Eq.~(\ref{WTI}) represents a fundamental bound on the exchange between energy and correlations, if the subsystems $S_{1}$ and $S_{2}$ are not interacting, that is, if the systems Hamiltonian is of the form $H\Sys=H\Sone+H\Stwo$. In this work we relax this assumption, and explore how the relation in Eq.~(\ref{WTI}) is modified for interacting systems.

An important difference to previous results lies in the fact that thermal states of interacting Hamiltonians are generally already correlated, and may potentially even be entangled~\cite{AmicoFazioOsterlohVedral2008}. This naturally raises the question of whether the presence of interactions provides advantages for the generation of (additional) correlations. We answer this question affirmatively, by constructing explicit strategies to achieve $W<T\Delta\mathcal{I}\Sonetwo$ for some energy range in any finite-dimensional system with arbitrary interacting Hamiltonian. While these procedures can improve on the best protocols for non-interacting systems, they are not necessarily optimal in the sense that other protocols may exist that generate more correlations at the same energy cost. To complement this approach we therefore develop optimal strategies for two physically relevant cases: two interacting, fermionic or bosonic modes.

This paper is structured as follows. We first provide a short summary of the framework for this investigation in Sec.~\ref{sec:framework}. We then approach the problem of generating correlations in interacting systems in Sec.~\ref{sec:results}, where we develop general strategies to use the energy contained in the interactions to improve upon the bound in Eq.~(\ref{WTI}), and explicitly demonstrate their applicability in a system of two qubits.  In Sec.~\ref{sec:two fermionic modes}, we then turn to another finite-dimensional example, two interacting, fermionic modes. This system, restricted by superselection rules, is amenable to a numerical approach that we use to determine the optimal conversion of energy into correlations. Finally, in Sec.~\ref{sec:two bosonic modes}, we study the generation of correlations in the infinite-dimensional system of two interacting, bosonic modes.

\section{Framework}\label{sec:framework}

We consider a bipartite system $S$, made up of  subsystems $S_{1}$ and $S_{2}$, initially at thermal equilibrium at ambient temperature  $T=1/\beta$, described by a thermal state
\vspace*{-1mm}
\begin{align}
    \tau(\beta)    &=\,\mathcal{Z}^{-1}(\beta)\,e^{-\beta H\Sys}\,,
    \label{eq:general thermal states}
    \vspace*{-2mm}
\end{align}
where $\mathcal{Z}(\beta)=\tr(e^{-\beta H\Sys})$ is the partition function, and $H\Sys$ is the system Hamiltonian. We further assume the presence of an auxiliary heat bath $B$, that is, an arbitrarily large system in thermal equilibrium with $S$. The total Hamiltonian is $H=H\Sys+H\bath\,$, and the initial state can be written as $\tau\SB(\beta)=\tau(\beta)\otimes\hspace*{0.5pt}\tau\bath(\beta)$. In order to transform this equilibrium state, we consider arbitrary unitary operations $U_{\rm SB}$ on $SB$. Since the joint systems $SB$ is closed, these unitaries correspond to the most general operations permissable in this situation. The average work cost  of transforming the state of $S$ from $\tau(\beta)$ to a final state $\rho=\tr\subA{B}(\USB\hspace*{0.5pt}\tau\SB(\beta)\USB^{\dagger})$ is given by,
\begin{align}
 W   &=\,\tr\Bigl(H\bigl[\USB\hspace*{0.5pt}\tau\SB(\beta)\USB^{\dagger}
    -\tau\SB(\beta)\bigr]\Bigr),
    \label{eq:W in}
\end{align}
which corresponds to the total external energy input (see, e.g., Refs.~\cite{PuszWoronowicz1978,Lenard1978}). In Ref.~\cite{EspositoVanDenBroeck2011} (see also Refs.~\cite{SkrzypczykShortPopescu2014,Aberg2013,ReebWolf2014} for the same result in related frameworks), it is shown that  $W$ can be bounded
by the (nonequilibrium) free energy difference,
\begin{align}
    W(\tau \rightarrow \rho)   &\geq\,\Delta F\Sys\,=\,F(\rho)-F(\tau),
    \label{eq:minimal energy cost}
\end{align}
where the free energy with respect to the reservoir at temperature~$T$ is
\begin{align}
    F(\rho) &=\,E(\rho)\,-\,T\hspace*{0.5pt}S(\rho)\,.
    \label{eq:non equilibrium free energy}
\end{align}
Here, $E(\rho)=\tr\bigl(H\Sys\rho\bigr)$ is the average energy, and $S(\rho)=-\tr\bigl(\rho \ln(\rho)\bigr)$ is the von Neumann entropy. Note that $F(\rho)$ depends only on the state of $S$ and the temperature of $B$. Equality in~(\ref{eq:minimal energy cost}) can be obtained in a quasistatic process~\cite{SkrzypczykShortPopescu2014,Aberg2013,ReebWolf2014}, in which case the work cost becomes minimal.

We now wish to invest some work~$W$ to increase the correlations within the bipartite state of $S$  as much as possible. (Note that $B$ is only an auxiliary system and we do not wish to create correlations between $S$ and $B$.) In Refs.~\cite{HuberPerarnauHovhannisyanSkrzypczykKloecklBrunnerAcin2015,BruschiPerarnauLlobetFriisHovhannisyanHuber2015} this problem was considered for noninteracting Hamiltonians, $H_S=H\Sone+H\Stwo$. Here, we want to depart from this paradigm and consider an interacting Hamiltonian of the form
\vspace*{-1mm}
\begin{align}
    H\Sys   &=\,H\Sone\,+\,H\Stwo\,+\,\HI\,.
    \label{eq:total Hamiltonian}
\end{align}
As discussed above, the work cost of transforming $\tau(\beta)$ to a final state $\rho$ satisfies, $W(\tau \rightarrow \rho)  \geq \Delta F\Sys\,=\,F(\rho)-F(\tau)$, and equality can be achieved in a quasi-static process and with  a sufficiently large bath \cite{SkrzypczykShortPopescu2014,Aberg2013,ReebWolf2014}. The task is then to maximize the correlations of $\rho$ under the constraint $F(\rho)-F(\tau)\leq W$, where $W$ is the amount of available work.

Before continuing, note that the main ingredients of the investigations in Refs.~\cite{HuberPerarnauHovhannisyanSkrzypczykKloecklBrunnerAcin2015,BruschiPerarnauLlobetFriisHovhannisyanHuber2015} are preserved:
\begin{enumerate}[\hspace*{-1mm}(i)]
\item{The initial state is in thermal equilibrium, and therefore, the energy cost of transforming $\tau(\beta)$ to any final state $\rho$ is nonnegative, $W\geq 0$.}
\vspace*{-1mm}
\item{We assume arbitrary (in particular, unitary) operations can be performed on $S$ and the auxiliary thermal bath, which allows obtaining fundamental bounds on the work cost of correlations.}
\end{enumerate}
We quantify the amount of correlations between the subsystems by the mutual information,
\begin{align}
    \mathcal{I}\Sonetwo(\rho) &=\,S(\rho\Sone)\,+\,S(\rho\Stwo)\,-\,S(\rho)\,,
    \label{eq:mutual inf definition}
\end{align}
where $\rho\Soneortwo=\tr\Stwoorone(\rho)$ are the reduced states of the subsystems. The main quantity of interest throughout this paper will be the correlations gain,
\begin{align}
    \Delta\mathcal{I}\Sonetwo   &=\,\mathcal{I}\Sonetwo(\rho)\,-\,\mathcal{I}\Sonetwo(\tau).
    \label{eq:mutual inf difference}
\end{align}
That is, we take a point of view inspired by Landauer's principle, and ask how many units of correlations $\Delta\mathcal{I}\Sonetwo$ can be newly generated (on top of the preexisting correlations) at the expense of one unit of energy. Note that, in the noninteracting case, $\Delta\mathcal{I}\Sonetwo$ and $\mathcal{I}\Sonetwo$ coincide, as the initial thermal state of the Hamiltonian $H\Sone+H\Stwo$ is an uncorrelated product state. In the interacting case, $\Delta\mathcal{I}\Sonetwo$ and $\mathcal{I}\Sonetwo$ still arise from the same optimization procedure, but $\mathcal{I}\Sonetwo\geq\Delta\mathcal{I}\Sonetwo$. Since $\Delta\mathcal{I}\Sonetwo$ quantifies the amount of correlations generated  through the investment of~$W$, we  first focus on $\Delta\mathcal{I}\Sonetwo$, establishing strategies to achieve $W<T\Delta\mathcal{I}\Sonetwo$ in Sec.~\ref{sec:results}.

\vspace*{-1mm}

\section{General Considerations}\label{sec:results}

\subsection{Work cost of generating correlations}

Let us now relate the correlation gain~$\Delta\mathcal{I}\Sonetwo$ to the minimal work cost~$\Delta F\Sys$. Inserting the Hamiltonian from Eq.~(\ref{eq:total Hamiltonian}) into~(\ref{eq:minimal energy cost}) one obtains
\begin{align}
    \Delta F\Sys    &=\,\tr\bigl(H\Sone[\rho-\tau]\,+\,H\Stwo[\rho-\tau]\bigr)\,+\,\tr\bigl(\HI[\rho-\tau]\bigr)\nonumber\\
    &\ + T\,\bigl[S(\tau)\,-\,S(\rho)\bigr]\,.
   \label{eq:minimal work cost rewritten}
\end{align}
On the other hand, for the difference in mutual information from Eq.~(\ref{eq:mutual inf difference}) one finds $T\,\Delta\mathcal{I}\Sonetwo=T\,\bigl[S(\tau)\,-\,S(\rho)\bigr]+T\,\bigl[S(\rho\Sone)-S(\tau\Sone)\bigr]+T\,\bigl[S(\rho\Stwo)-S(\tau\Stwo)\bigr].$ After some straightforward manipulations we then arrive at
\begin{align}
    \Delta F\Sys    &=T\Delta\mathcal{I}\Sonetwo+\tr\bigl(\HI[\rho-\tau]\bigr)+\Delta\tilde{F}\Sone+\Delta\tilde{F}\Stwo\,,
    \label{eq:Delta FSys}
\end{align}
where the quantities $\tilde{F}\Soneortwo$ correspond to nonequilibrium free energies with respect to the local Hamiltonians, i.e.,
\begin{align}
    \tilde{F}\Si (\rho)   &=\,\tr\bigl(H\Si \rho\bigr)\,-\,T S(\rho\Si)
    \label{eq:nonint free energy change Sone and Stwo}
\end{align}
with $\rho\Soneortwo= \tr_{\Stwoorone} \rho$. Here, it is important to note that the marginals $\tau\Soneortwo= \tr_{\Stwoorone} \tau$ of the initial thermal state are not themselves thermal states with respect to the local Hamiltonians, $\tau\Soneortwo\neq\mathcal{Z}\Si^{-1}\exp(-\beta H\Si)$. It can therefore be inferred that, while $\Delta F\Sys\geq0$ since the initial state $\tau$ is at thermal equilibrium with the bath, the quantities $\Delta\tilde{F}\Si$ may have either sign. Only when $\HI$ vanishes are the marginals of~$\tau$ also thermal states with minimal free energy and $\sum_{i}\Delta\tilde{F}\Si$ is nonnegative. In this case one obtains (see Ref.~\cite{BruschiPerarnauLlobetFriisHovhannisyanHuber2015} for a detailed derivation) the bound
\begin{align}
T \Delta \hspace*{1pt}\mathcal{I}\Sonetwo(\rho)\leq \Delta F\Sys \leq W, \hspace{5mm} {\rm if} \hspace{2mm} \HI=0 ,
\label{eq: non-interacting bound}
\end{align}
which shows that, for noninteracting systems, at least $T\Delta \hspace*{1pt}\mathcal{I}\Sonetwo$ units of work have to be invested to increase the correlations of the systems by the amount $\Delta \hspace*{1pt}\mathcal{I}\Sonetwo$.
\vspace*{-1mm}

\subsection{Strategies to utilize the interactions}\label{sec:Strategies to utilize interactions to generate correlations}

In this section we determine strategies that can potentially outperform the bound of Eq.~(\ref{eq: non-interacting bound}) by making use of the energy provided by the interactions. We formulate these strategies for arbitrary Hamiltonians of bipartite systems, whose subsystems $S_{1}$ and $S_{2}$ can be arbitrary finite-dimensional quantum systems, qudits, with Hilbert spaces $\mathcal{H}\Sone$ and $\mathcal{H}\Stwo$, and dimensions $d_{1}$ and $d_{2}$, respectively. The density operators for such systems can be written in~a generalized Bloch-Fano decomposition~\cite{Fano1983,BertlmannKrammer2008b}, that is
\vspace*{-1mm}
\begin{align}
    \rho    &=\,\frac{1}{d_{1}d_{2}}\Bigl(\mathds{1}\Sys\,+\,\sum\limits_{m=1}^{d_{1}^{2}-1}a_{m}\,\sigone_{m}\otimes\mathds{1}\Stwo
        +\,\sum\limits_{n=1}^{d_{2}^{2}-1}b_{n}\mathds{1}\Sone\!\otimes\sigtwo_{n}\nonumber\\
        &\qquad +\,\sum\limits_{m=1}^{d_{1}^{2}-1}\sum\limits_{n=1}^{d_{2}^{2}-1}\,c_{mn}\,\sigone_{m}\otimes\sigtwo_{n}\Bigr)\,,
    \label{eq:density operator bloch decomp}
\end{align}
where the Hermitean operators $\sigi_{m}$ satisfy $\tr(\sigi_{m}\sigi_{n})=2\delta_{mn}$ and $\tr(\sigi_{m})=0$, and the real coefficients $a_{m}$, $b_{n}$, and $t_{mn}$ are subject to constraints arising from the positivity of~$\rho$. The reduced states are then immediately obtained as
\vspace*{-1mm}
\begin{subequations}
\label{eq:red state bloch decomp}
    \begin{align}
        \rho\Sone   &=\,\frac{1}{d_{1}}\Bigl(\mathds{1}\Sone\,+\,\sum\limits_{m=1}^{d_{1}^{2}-1}a_{m}\,\sigone_{m}\Bigr)\,,
            \label{eq:red state bloch decomp S1}\\
        \rho\Stwo   &=\,\frac{1}{d_{2}}\Bigl(\mathds{1}\Stwo\,+\,\sum\limits_{n=1}^{d_{2}^{2}-1}b_{n}\,\sigtwo_{n}\Bigr)\,.
            \label{eq:red state bloch decomp S2}
    \end{align}
\end{subequations}
The Hermitean interaction Hamiltonian can similarly be written as
\begin{align}
    \HI &=\,\sum\limits_{m=1}^{d_{1}^{2}-1}\sum\limits_{n=1}^{d_{2}^{2}-1}\,\epsilon_{mn}\,\sigone_{m}\otimes\sigtwo_{n}\,,
    \label{eq:Bloch fano interaction term}
\end{align}
with real coefficients $\epsilon_{mn}$. Any terms of the form $\mathds{1}\Sone\otimes\sigtwo_{m}$ and $\sigone_{m}\otimes\mathds{1}\Stwo$ that may appear in such~a decomposition of~$\HI$ can be absorbed into the local Hamiltonians $H\Si$.

Returning to the relation of Eq.~(\ref{eq:Delta FSys}), notice that the interactions allow surpassing the bound in Eq.~(\ref{eq: non-interacting bound}) whenever $\tr\bigl(\HI[\rho-\tau]\bigr)+\sum_{i}\Delta\tilde{F}\Si$ is negative. The expansion of Eq.~(\ref{eq:density operator bloch decomp}) further permits treating each of these terms independently: The terms $\Delta\tilde{F}\Si$ depend only on the local Bloch vector components $a_{m}$ and $b_{n}$, for $i=1$ and $i=2$, respectively, whereas the interaction term $\tr\bigl(\HI[\rho-\tau]\bigr)$ depends only on the correlation tensor $c_{mn}$. With this we can formulate two complementary strategies. However, it is important to keep in mind that any choice of the coefficients $c_{mn}$, $a_{m}$ and $b_{n}$ is subject to the positivity constraint $\rho\geq0$.

First, we focus on the local terms $\Delta\tilde{F}\Si$. Defining the local Gibbs states as $\gamma\Si\equiv \mathcal{Z}\Si^{-1} e^{-\beta H\Si}$, which are generically different from the local initial states $\tau\Soneortwo=\tr_{\Stwoorone} \tau$, it is useful to rewrite $\Delta\tilde{F}\Si$ as,
\begin{align}
    \beta\Delta\tilde{F}\Si &=\,\beta\bigl(F(\rho\Si)\,-\, F(\gamma\Si)\bigr)\,-\,\beta\bigl(F(\tau\Si)\,-\,F(\gamma\Si)\bigr)\nonumber\\[1mm]
    &=\,S(\rho\Si\dvline\gamma\Si)\,-\,S(\tau\Si\dvline\gamma\Si)\,,
\label{eq: DeltaTildeF}
\end{align}
where $S(\rho\dvline\tau)=-S(\rho)-\tr(\rho\ln\tau)$ is the relative entropy. Here, we have used that $\beta\bigl[F(\rho)-F\bigl(\tau(\beta)\bigr)\bigr]=S(\rho\dvline\tau(\beta))$, which can easily be shown using Eqs.~(\ref{eq:general thermal states}) and~(\ref{eq:non equilibrium free energy}). Since $S(\,.\,\dvline\,.\,)$ is a measure of distance between two quantum states, the quantities $\Delta\tilde{F}\Si$ are negative whenever the final reduced states $\rho\Si$ are closer to the local Gibbs states $\gamma\Si$ than the initial state marginals $\tau\Si$. This provides a simple strategy to minimize $\Delta\tilde{F}\Si$: The Bloch coefficients $a_{m}^{(\rho)}$ and $b_{n}^{(\rho)}$ of the final state $\rho$ should to be chosen as close as possible to $a_{m}^{(\gamma)}$ and $b_m^{(\gamma)}$, respectively, where $a_{m}^{(\gamma)}=\tfrac{d_{1}}{2}\tr(\gamma\Sone\sigone_{m})$ and $b_{n}^{(\gamma)}=\tfrac{d_{2}}{2}\tr(\gamma\Stwo\sigtwo_{n})$. This strategy ensures that $\Delta \tilde{F}\Si <0 $.

The second strategy entails the minimization of the term $\tr\bigl(\HI[\rho-\tau]\bigr)$. Using Eqs.~(\ref{eq:density operator bloch decomp}) and~(\ref{eq:Bloch fano interaction term}), we can express it in terms of the correlation tensors $c_{mn}^{(\rho)}$ and $c_{mn}^{(\tau)}$ of $\rho$ and $\tau$, respectively, obtaining
\begin{align}
    \tr\bigl(\HI [\rho-\tau]\bigr) &= \,\sum\limits_{m=1}^{d_{1}^{2}-1}\sum\limits_{n=1}^{d_{2}^{2}-1}\,\bigl(c_{mn}^{(\rho)}-c_{mn}^{(\tau)}\bigr)\epsilon_{mn}\,.
    \label{eq:corr tensor condition}
\end{align}
This relation has~a clear geometrical interpretation. Mapping $c_{mn}^{(\rho)}$, $c_{mn}^{(\tau)}$, and $\epsilon_{mn}$ to vectors $\bf{c}^{(\rho)}$, $\bf{c}^{(\tau)}$, and ${\boldsymbol \epsilon}$ in~a Euclidean vector space of dimension $(d_{1}^{2}-1)(d_{2}^{2}-1)$, the condition of Eq.~(\ref{eq:corr tensor condition}) becomes
\begin{align}
\tr\bigl(\HI [\rho-\tau]\bigr) &=\,(\bf{c}^{(\rho)}-\bf{c}^{(\tau)})\cdot \boldsymbol{\epsilon}\,.
    \label{eq:corr tensor condition 2}
\end{align}
To minimize the expression in~(\ref{eq:corr tensor condition 2}) it is hence desirable to select the vector ($\bf{c}^{(\rho)}-\bf{c}^{(\tau)}$) to be as antiparallel as possible to $\boldsymbol{\epsilon}$.

The considerations discussed in this section hence provide two complementary strategies to obtain $\beta W< \Delta\mathcal{I}\Sonetwo $, as desired. In general, the choices of $a_m^{(\rho)}$, $b_m^{(\rho)}$, and $c_{mn}^{(\rho)}$ are limited by the positivity constraint, $\rho \geq 0$ (and of course also by the amount of available work, $W$). In the next section we illustrate possible issues with the positivity of $\rho$ in more detail for a particular example of two interacting qubits.

\subsection{Improved generation of correlations for two qubits}\label{sec:Improved generation of correlations for two qubits}

We consider a system of two qubits, coupled by the Hamiltonian
\begin{align}
    H\Sys   &=\,\omega\,\bigl(\sigone_{z}\,+\,\sigtwo_{z}\bigr)\,+\,\epsilon\,\sigone_{z}\!\otimes\sigtwo_{z}\,,
    \label{eq:two qubit system Hamiltonian}
\end{align}
where $\omega\geq0$ and $\epsilon\in\mathbb{R}$ can take either sign. In this simple example, the presence of the interaction Hamiltonian $\HI=\epsilon\,\sigone_{z}\!\otimes\sigtwo_{z}$ does not change the eigenstates of~$H\Sys$, but the eigenvalues of the noninteracting system are modified to $(\epsilon\pm2\omega)$ and $-\epsilon$ (twice degenerate). The initial thermal state $\tau(\beta)=e^{-\beta H\Sys}/\mathcal{Z}$ is hence of the form
\begin{align}
    \tau(\beta) &=\,\mathcal{Z}^{-1}
    \diag\{e^{-\beta(\epsilon+2\omega)},e^{\beta\epsilon},e^{\beta\epsilon},e^{-\beta(\epsilon-2\omega)}\}
    \label{eq:two qubits thermal state}
\end{align}
with $\mathcal{Z}=\tr\bigl(e^{-\beta H\Sys}\bigr)\geq0$. The nonzero coefficients of the Bloch decomposition of $\tau(\beta)$ are
\begin{subequations}
\begin{align}
    a_{z}^{(\tau)}  &=\,b_{z}^{(\tau)}\,=\,-\,2\mathcal{Z}^{-1}e^{-\beta\epsilon}\sinh(2\beta\omega)\,<\,0\,,
    \label{eq:two qubits bloch coefficient az}\\
    c_{zz}^{(\tau)} &=\,1\,-\,\frac{4e^{\beta\epsilon}}{\mathcal{Z}}\,.
    \label{eq:two qubits bloch coefficient czz}
\end{align}
\end{subequations}

To correlate the system, we apply a two-step protocol based on the strategies discussed in Sec.~\ref{sec:Strategies to utilize interactions to generate correlations}. In the first phase of the protocol, step~$\mathrm{I}$, we aim to minimize the term $\tr\bigl(\HI [\rho-\tau]\bigr)$. To do this, we transform the state $\tau$ to $\rhoI$, such that the local Bloch vector components remain invariant, $a_{z}\suprhoI=b_{z}\suprhoI=a_{z}^{(\tau)}$, while the (nonzero) correlation tensor coefficient is mapped to
\begin{align}
    c_{zz}\suprhoI    &=\,c_{zz}^{(\tau)}\,-\,\operatorname{sgn}(\epsilon)\,\alphaI\,,
    \label{eq:step I correlation tensor}
\end{align}
for $\alphaI\geq0$. With this, one finds $\tr\bigl(\HI [\rhoI-\tau]\bigr)=-|\epsilon|\alphaI$ and from Eq.~(\ref{eq:Delta FSys}) we obtain
\begin{align}
    W_{\mathrm{I}}  &=\,T\hspace*{1pt}\Delta\mathcal{I}\Sonetwo\,-\,|\epsilon|\alphaI\,,
    \label{eq:two qubits step one work cost}
\end{align}
where we assumed that the process is quasistatic, so that $W=\Delta F\Sys$. (Note that the same assumption is made also later in step~$\mathrm{I\hspace*{-0.5pt}I}$.)
The correlations are hence generated at a work cost that is lower than in the noninteracting case, $W_{\mathrm{I}}\leq T\hspace*{1pt}\Delta\mathcal{I}\Sonetwo$. However, it is crucial to note that the transformation in Eq.~(\ref{eq:step I correlation tensor}) is limited by the positivity constraint, $\rho_I \geq 0$, requiring $2|a_{z}^{(\tau)}|-1\leq c_{zz}\suprhoI\leq1$. Depending on the sign of the interaction term, one of these bounds is reached, when enough energy is supplied. That is, $c_{zz}\suprhoI$ eventually tends towards either $c_{zz}\suprhoI=2|a_{z}^{(\tau)}|-1$ or $c_{zz}\suprhoI=1$ for $\epsilon>0$ or $\epsilon<0$, respectively.

\begin{figure}[ht!]
\label{fig:two qubits epsilon smaller 0 Delta F tilde}
\includegraphics[width=0.45\textwidth,trim={0cm 0mm 0cm 0mm}]{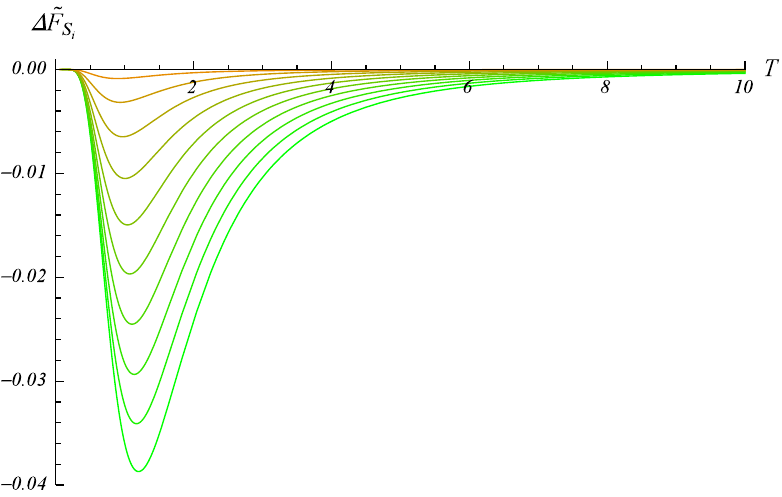}
\caption{
\textbf{Advantage in correlation cost:} During step~$\mathrm{I\hspace*{-0.5pt}I}$ of the protocol to generate correlations between two qubits, an advantage over the noninteracting case arises when $\Delta\tilde{F}\Si$ from Eq.~(\ref{eq:nonint free energy change Sone and Stwo}) becomes negative. $\Delta\tilde{F}\Si$ is plotted here against the temperature~$T$ in units of $\omega$ (recall that we use units where $\hbar=k_{\protect\raisebox{-0pt}{\tiny{B}}}=1$) for $\alphaII=0.5$, and the different curves correspond to values of $\epsilon$ (also in units of $\omega$) from $\epsilon=0$ (top) to $\epsilon=-1$ (bottom) in steps of $0.1$. The advantage increases with increasing coupling strength~$\epsilon$, but does not monotonically decrease with the temperature. Instead, the advantage becomes maximal at a finite temperature. Although curves are only shown for a fixed value $\alphaII=0.5$, we have checked that other values yield analogous behaviour and the advantage increases monotonically with~$\alphaII$.}
\end{figure}

If more energy is available than is needed to saturate the positivity constrain in step~$\mathrm{I}$, we may employ the complementary strategy discussed in Sec.~\ref{sec:Strategies to utilize interactions to generate correlations} in step~$\mathrm{I\hspace*{-0.5pt}I}$, the second phase of the protocol. Now, we keep the correlation tensor fixed, while changing the local Bloch vector components to minimize $\Delta\tilde{F}\Si$. This entails moving the marginals closer to the states $\gamma\Si$ that are locally thermal with respect to $H\Si$. These local Gibbs states are here given by
\begin{align}
    \gamma\Si &=\,\frac{e^{-\beta H\Si}}{\mathcal{Z}\Si}\,=\,\tfrac{1}{2}\bigl(\idN{2}\,-\,\tanh(\beta\omega)\sigi_{z}\bigr)\,,
    \label{eq:two quibts locally thermal states}
\end{align}
with $a_{z}^{(\gamma)}=-\tanh(\beta\omega)<0$. We hence map $\rhoI$ to the state $\rhoII$ with Bloch vector components given by
\begin{align}
    a_{z}\suprhoII  &=\,(1-\alphaII)\,a_{z}^{(\tau)}\,+\,\alphaII\,a_{z}^{(\gamma)}\,,
    \label{eq:step II bloch vector components}
\end{align}
where $0\leq\alphaII\leq1$. Again, the positivity constraint $\rhoII\geq0$ must still be taken into account. For $\epsilon<0$ we find that the full range of $\alphaII$ is compatible with the positivity of $\rhoII$. The work cost of step~$\mathrm{I\hspace*{-0.5pt}I}$ is given by $W_{\mathrm{I\hspace*{-0.5pt}I}}=  T\Delta\mathcal{I}\Sonetwo+\Delta\tilde{F}\Sone+\Delta\tilde{F}\Stwo\,,$ and, as illustrated in Fig.~\ref{fig:two qubits epsilon smaller 0 Delta F tilde}, we indeed find that $\Delta\tilde{F}\Si\leq0$ for all values of~$T\geq0$, $0\leq\alphaII\leq1$, and $\epsilon<0$.

For $\epsilon>0$, on the other hand, the positivity constraints require that $|a_{z}\suprhoII|\leq|a-{z}^{(\tau)}|$. Since $a_{z}^{(\tau)}<0$ and $a_{z}^{(\gamma)}=-\tanh(\beta\omega)<0$, Eq.~(\ref{eq:step II bloch vector components}) yields $|a_{z}\suprhoII|=(1-\alphaII)|a_{z}^{(\tau)}|+\alphaII\,\tanh(\beta\omega)\geq|a_{z}^{(\tau)}|$. Unfortunately, since $|a_{z}^{(\tau)}|=\sinh(2\beta\omega)/\bigl(\cosh(2\beta\omega)+e^{-\beta\epsilon}\bigr)\leq\tanh(\beta\omega)$, one finds that $|a_{z}\suprhoII|\geq|a-{z}^{(\tau)}|$, that is, the positivity constraint does not allow for step~$\mathrm{I\hspace*{-0.5pt}I}$ of the protocol to be carried out for $\epsilon>0$.

In addition to the strategies discussed here, the states obtained after steps~$\mathrm{I}$ and $\mathrm{I\hspace*{-0.5pt}I}$ may be further correlated until the maximal value of correlation is reached. However, the work cost per newly generated unit of correlation beyond this point may be the same, or even higher than in the noninteracting case. We shall shed light on this possibility in the next sections, by studying in detail protocols to correlate bipartite systems of fermions and bosons.

\section{Two fermionic modes}\label{sec:two fermionic modes}

In this section, we consider the two systems $S_{1}$ and $S_{2}$, which are to be correlated, to be two fermionic modes with ladder operators $\bone$, $\bonedg$ and $\btwo$, $\btwodg$, respectively. Such fermionic systems have been well studied in the context of quantum information processing, see, e.g., Refs.~\cite{CabanPodlaskiRembielinskiSmolinskiWalczak2005,BanulsCiracWolf2007,BalachandranGovindarajanDeQueirozReyesLega2013}, but we shall briefly review some key features. The annihilation and creation operators satisfy the usual anticommutation relations $\{\bn{m},\bndg{n}\}=\delta_{mn}$ and $\{\bn{m},\bn{n}\}=0$. While the Pauli exclusion principle limits the dimension of the corresponding two-mode Fock space to~$4$, the anticommutation relations nonetheless imply~a different subsystem structure as compared to~a two-qubit Hilbert space~\cite{FriisLeeBruschi2013}. Despite this inequivalence\footnote{Note that $n$-mode fermionic Fock spaces are isomorphic to $n$-qubit Hilbert spaces via maps such as the Jordan-Wigner transformation. However, local operators in one space are generally mapped to nonlocal ones in the other. The marginals of an $n$-mode fermionic state~$\rho$ are hence generally not isomorphic to those of the $n$-qubit state~$\tilde{\rho}$, even if $\tilde{\rho}$ is related to~$\rho$ via an isomorphism.} of qubits and fermionic modes, the marginals and correlation measures for the fermionic system are well-defined when imposing~a superselection rule that forbids superpositions of even and odd numbers of fermions~\cite{Friis2016,AmosovFilippov2015}.

The fermionic system is hence of interest for the following reasons: First, the anticommutation relations and the restrictions of the superselection rule provide~a qualitative difference to the qubit case, which makes for an interesting comparison. Second, the low-dimensional Hilbert space is amenable to a numerical treatment, allowing~a rather general approach to optimal protocols for the generation of correlations. Third, fermionic fields form~a conceptually fundamental ingredient in the current view of matter in the universe in terms of relativistic quantum field theory.

\subsection{Hamiltonian and initial thermal state}\label{sec:fermionic thermal state}

Let us now turn to the specific system Hamiltonian that we consider in this section, $H=H\Sone\,+\,H\Stwo+\HI$.  For the noninteracting part, we consider the standard Hamiltonian for two modes of the same frequency $\omega$, i.e.,
\begin{align}
    H\Sone\,+\,H\Stwo   &=\,\omega\,\bigl(\bonedg\bone\,+\,\btwodg\btwo\bigr)\,.
    \label{eq:fermionic noninteracting hamiltonian}
\end{align}
For the interaction between the modes, we will employ the most general two-mode coupling term that is quadratic in the mode operators, given by
\begin{align}
    \HI &=\,H_{\mathrm{even}}\,+\,H_{\mathrm{odd}}\nonumber\\[1mm]
    &\ =\,\epsilon_{\mathrm{even}}\bigl(\bone\btwo\,+\,\btwodg\bonedg\bigr)
        \,+\,\epsilon_{\mathrm{odd}}\bigl(\bonedg\btwo\,+\,\btwodg\bone\bigr)\,,
    \label{eq:fermionic interacting hamiltonian}
\end{align}
where $H_{\mathrm{even}}$ couples only the states $\fket{0}$ and $\fket{1_{1}}\fket{1_{2}}=\bonedg\btwodg\fket{0}$ in the even subspace, while $H_{\mathrm{odd}}$ acts in the odd subspace spanned by $\fket{1_{1}}=\bonedg\fket{0}$ and $\fket{1_{2}}=\btwodg\fket{0}$. Here $\fket{0}$ is the vacuum state satisfying $\bn{i}\fket{0}=0\ \forall~i$, and the double-lined ket notation indicates the antisymmetrized tensor product for the excited states, i.e., $\fket{1_{1}}\fket{1_{2}}=\fket{1_{1}}\wedge\fket{1_{2}}=-\fket{1_{2}}\fket{1_{1}}$ (see, e.g., Ref.~\cite{FriisLeeBruschi2013} for more information).

The thermal state $\tau(\beta)$ of $H$ can be computed straightforwardly. The eigenvalues of $H$ read,
\begin{subequations}
\label{eq:fermion eigenvalues}
\begin{align}
    \lambda_{1,4}   &=\,\omega\,\pm\,\sqrt{\omega^{2}+\epsilon_{\mathrm{even}}^{2}}\,,
    \label{eq:fermion eigenvalues even}\\[1mm]
    \lambda_{2,3}   &=\,\omega\,\pm\,\epsilon_{\mathrm{odd}}\,,
    \label{eq:fermion eigenvalues odd}
\end{align}
\end{subequations}
where the labels $3$, $4$ refer to the negative relative sign, and the corresponding eigenstates are given by
\begin{subequations}
\label{eq:fermion eigenstates}
\begin{align}
    \fket{\lambda_{1,4}}   &=\,\frac{1}{\sqrt{\epsilon_{\mathrm{even}}^{2}+\lambda_{1,4}^{2}}}\Bigl(\epsilon_{\mathrm{even}}\fket{0}-\lambda_{1,4}\fket{1_{1}}\!\fket{1_{2}}\Bigr)\,,
    \label{eq:fermion eigenstates even}\\
    \fket{\lambda_{2,3}}   &=\,\frac{1}{\sqrt{2}}\Bigl(\fket{1_{2}}\,\pm\,\fket{1_{1}}\Bigr)\,.
    \label{eq:fermion eigenstates odd}
\end{align}
\end{subequations}
Then, $\tau(\beta)$ can the be written as,
\begin{align}
    \tau(\beta) &=\,\mathcal{Z}^{-1}(\beta)\sum\limits_{i}e^{-\beta\lambda_{i}}\,\fket{\lambda_{i}}\!\fbra{\lambda_{i}}\,,
    \label{eq:fermionic thermal state}
\end{align}
where the partition function is $\mathcal{Z}(\beta)=\sum_{i}e^{-\beta\lambda_{i}}$. It is important to note that $\tau(\beta)$ already contains correlations, which are computed in detail in the~\hyperref[sec:CorrelationsFermions]{Appendix}.

\subsection{Generation of Correlations}\label{sec:fermion numerics}

We now consider the task of correlating $\tau(\beta)$ further. The simple structure of the system (an only four-dimensional Hilbert space that is further restricted by superselection rules) allows us to consider the most general protocols beyond the strategies discussed in Sec.~\ref{sec:Strategies to utilize interactions to generate correlations}. That is, given some available work $W$, we consider the possibility to transform $\tau$ to any state $\rho$, provided $\Delta F\Sys(\tau\rightarrow\rho)\leq W$ is satisfied. In order to maximize the created correlations, $\Delta\mathcal{I}\Sonetwo$, we conveniently parametrize the final state~$\rho$, and numerically optimize its mutual information $\mathcal{I}\Sonetwo(\rho)$ under the constraint of~a maximally available free energy.

\begin{figure}[ht!]
\label{fig:fermionic newly generated correlation}
(a)\hspace*{-1mm}\includegraphics[width=0.47\textwidth]{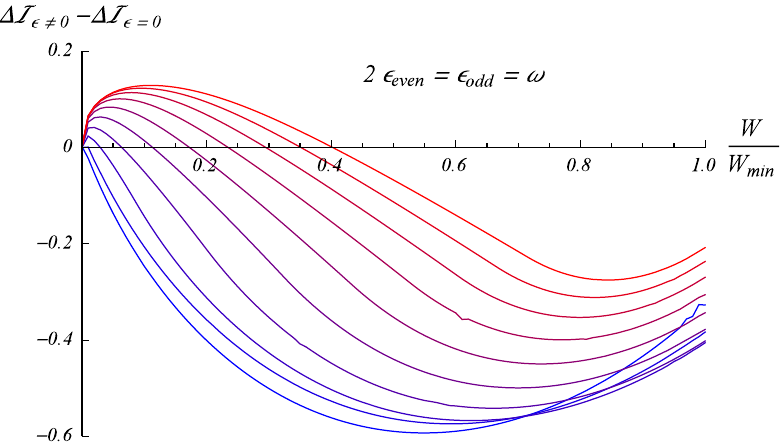}
\vspace*{0.5mm}
(b)\hspace*{-1mm}\includegraphics[width=0.47\textwidth]{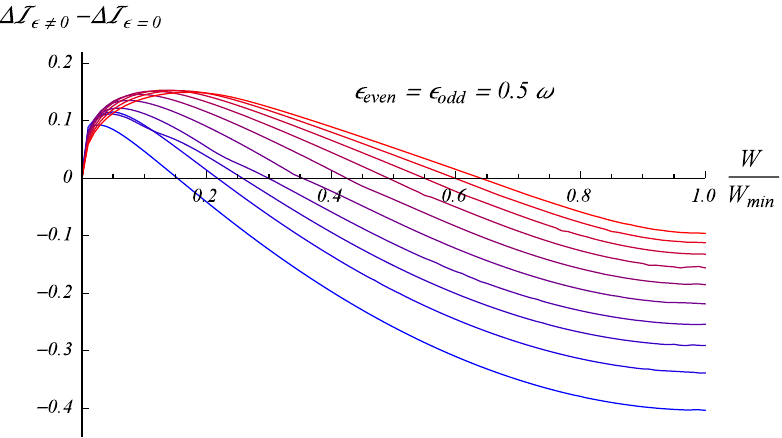}
\vspace*{0.5mm}
(c)\hspace*{-1mm}\includegraphics[width=0.47\textwidth]{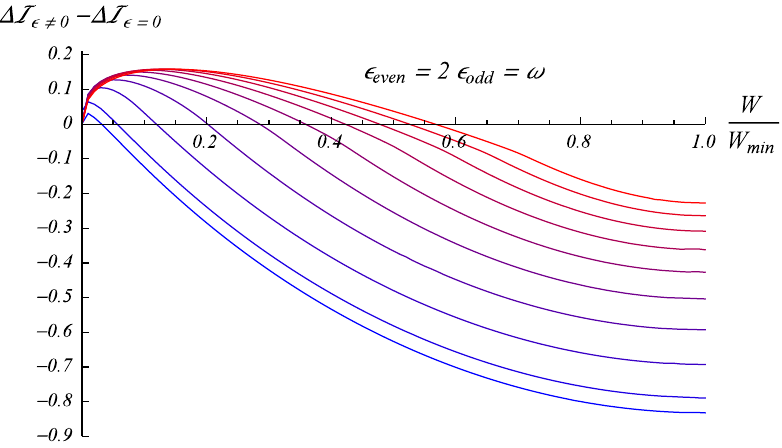}
\caption{
\textbf{Fermionic newly generated correlation cost:} The difference in correlations that can be newly generated for an available energy $W$ (in units of $W_{\mathrm{min}}$) in the presence ($\Delta\mathcal{I}_{\epsilon\neq0}$) and absence ($\Delta\mathcal{I}_{\epsilon=0}$) of interactions is shown for temperatures $T=0.1,\ldots,1$ (in units of $\hbar\omega/k_{B}$) in steps of $0.1$ (blue to red, top to bottom) for the ratios $\epsilon_{\mathrm{even}}/\epsilon_{\mathrm{odd}}=2,1$, and $0.5$ in (a),(b), and (c), respectively.}
\end{figure}

Since the final state needs to respect the superselection rule that forbids superpositions of even and odd numbers of fermions~\cite{Friis2016}, the four-dimensional Fock space splits into two two-dimensional spaces. An arbitrary two-mode final state may therefore be written as a convex combination of two density operators, $\rho_{\mathrm{even}}$ and $\rho_{\mathrm{odd}}$, corresponding to the subspaces of even and odd fermion numbers, respectively. We hence write
\begin{align}
    \rho    &=\,p\,\rho_{\mathrm{even}}\,+\,(1-p)\,\rho_{\mathrm{odd}}\,,
    \label{eq:fermion final state parametrization}
\end{align}
where $0<p<1$\,. For each of the two subspaces, we then use~a single-qubit Bloch representation, i.e.,
\begin{subequations}
\label{eq:fermion final state bloch rep subspaces}
\begin{align}
    \rho_{\mathrm{even}}    &=\,\tfrac{1}{2}\Bigl(\bigl[1+z_{\mathrm{even}}\bigr]\fket{\!0\!}\!\fbra{\!0\!}\label{eq:fermion final state bloch rep subspace even}\\
    &\ \ \ +\,\bigl[1-z_{\mathrm{even}}\bigr]\fket{\!1_{1}\!}\!\fket{\!1_{2}\!}\!\fbra{\!1_{2}\!}\!\fbra{\!1_{1}\!}
    \nonumber\\
    &\ \ \ +\,\bigl[(x_{\mathrm{even}}-iy_{\mathrm{even}})\fket{\!0\!}\!\fbra{\!1_{2}\!}\!\fbra{\!1_{1}\!}\,+\,\mathrm{H.~c.}\,\bigr]\Bigr)\,,
    \nonumber\\[1mm]
    \rho_{\mathrm{odd}} &=\,\tfrac{1}{2}\Bigl(\bigl[1+z_{\mathrm{odd}}\bigr]\fket{\!1_{2}}\!\fbra{\!1_{2}\!}\,+\,\bigl[1-z_{\mathrm{odd}}\bigr]\fket{\!1_{1}}\!\fbra{\!1_{1}\!}
    \nonumber\\
    &\ \ \ +\,\bigl[(x_{\mathrm{odd}}-iy_{\mathrm{odd}})\fket{\!1_{2}\!}\!\fbra{\!1_{1}\!}\,+\,\mathrm{H.~c.}\,\bigr]\Bigr)\,,
    \label{eq:fermion final state bloch rep subspace odd}
\end{align}
\end{subequations}
where the coefficients satisfy $|x_{\mathrm{even,odd}}|\leq1$, $|y_{\mathrm{even,odd}}|\leq1$, $|z_{\mathrm{even,odd}}|\leq1$, and
\begin{subequations}
\label{eq:fermion final state bloch rep constraints}
\begin{align}
    r^{2}_{\mathrm{even}}   &=\,x^{2}_{\mathrm{even}}+y^{2}_{\mathrm{even}}+z^{2}_{\mathrm{even}}\,\leq\,1\,,\\[1mm]
    r^{2}_{\mathrm{odd}}    &=\,x^{2}_{\mathrm{odd}}+y^{2}_{\mathrm{odd}}+z^{2}_{\mathrm{odd}}\,\leq\,1\,.
\end{align}
\end{subequations}
In this parametrization, the entropy of the final state can easily be obtained via its eigenvalues $\tfrac{p}{2}(1\pm r_{\mathrm{even}})$ and $\tfrac{1-p}{2}(1\pm r_{\mathrm{odd}})$. The energy of the final state, in turn, is
\begin{align}
    E(\rho) &=\omega (1 - p\,z_{\mathrm{even}})-p\,\epsilon_{\mathrm{even}} x_{\mathrm{even}}+(1 - p)\epsilon_{\mathrm{odd}} x_{\mathrm{odd}}.
    \label{eq:fermion final state energy}
\end{align}
Lastly, the final state marginals are of the form
\begin{subequations}
\label{eq:fermion final state marginals}
\begin{align}
    \rho\Sone   &=\,\tfrac{1}{2}\bigl(1+p\,z_{\mathrm{even}}+(1-p)z_{\mathrm{odd}}\bigr)\fket{\!0\!}\!\fbra{\!0\!}\label{eq:fermion final state marginal 1}\\[1mm]
    &\ \ \ +\tfrac{1}{2}\bigl(1-p\,z_{\mathrm{even}}-(1-p)z_{\mathrm{odd}}\bigr)\fket{\!1_{1}\!}\!\fbra{\!1_{1}\!}\,,
    \nonumber\\[1.5mm]
    \rho\Stwo   &=\,\tfrac{1}{2}\bigl(1+p\,z_{\mathrm{even}}-(1-p)z_{\mathrm{odd}}\bigr)\fket{\!0\!}\!\fbra{\!0\!}\label{eq:fermion final state marginal 2}\\[1mm]
    &\ \ \ +\tfrac{1}{2}\bigl(1-p\,z_{\mathrm{even}}+(1-p)z_{\mathrm{odd}}\bigr)\fket{\!1_{2}\!}\!\fbra{\!1_{2}\!}\,.
    \nonumber
\end{align}
\end{subequations}
For the illustration of the results, it is convenient to specify the amount of available input energy in units of~$W_{\mathrm{min}}$, the minimal free energy difference\footnote{Note that the minimal energy for maximal correlations ($W_{\mathrm{min}}$) depends on the coupling strength. Therefore, the functions $\Delta\mathcal{I}_{\epsilon\neq0}$ and $\Delta\mathcal{I}_{\epsilon=0}$, whose difference is plotted in Fig.~\ref{fig:fermionic newly generated correlation}, would be multiplied by different values when converting the plots of Fig.~\ref{fig:fermionic newly generated correlation} to absolute energy costs. This would result in shifted intersections with the horizontal axes. Nonetheless, since $W_{\mathrm{min}}$ is maximal in the absence of interactions, the intersections would all shift to the left, leaving the conclusion unchanged, that the presence of interactions may make the creation of new correlations more expensive, even if the overall amount of correlations is larger in the end.} to~a maximally correlated state. Taking into account that such~a maximally correlated state must be pure, and that the free energy of the initial state is $F(\tau)=-T\ln(\mathcal{Z})$, we find
\begin{align}
    W_{\mathrm{min}}    &=\,\omega\,-\,\max\{|\epsilon_{\mathrm{even}}|,|\epsilon_{\mathrm{odd}}|\}\,+\,T\,\ln(\mathcal{Z})\,.
    \label{eq:W min}
\end{align}
With this, we may numerically evaluate the maximal amount of correlations that can be created at~a fixed temperature~$T$ and~a fixed energy input~$W/W_{\mathrm{min}}$. The results of the optimization allows us to compare the energy cost of optimal protocols to generate correlations, for both interacting ($\epsilon_{\mathrm{even,odd}}\neq0$) and noninteracting systems ($\epsilon_{\mathrm{even,odd}}=0$). The results are shown in Fig.~\ref{fig:fermionic newly generated correlation}.

In agreement with our considerations in Sec.~\ref{sec:Strategies to utilize interactions to generate correlations}, one observes an initial regime where the interactions provide an advantage. However, at some point, the energy cost of $\Delta\mathcal{I}\Sonetwo$ becomes higher for interacting systems than for noninteracting ones. This is to be expected: Since the interacting system is correlated initially, the maximal value of $\Delta\mathcal{I}\Sonetwo$ is always lower than in the noninteracting case.

\begin{figure}[ht!]
\label{fig:fermionic generated correlation}
(a)\includegraphics[width=0.445\textwidth,trim={0cm 0mm 0cm 2mm}]{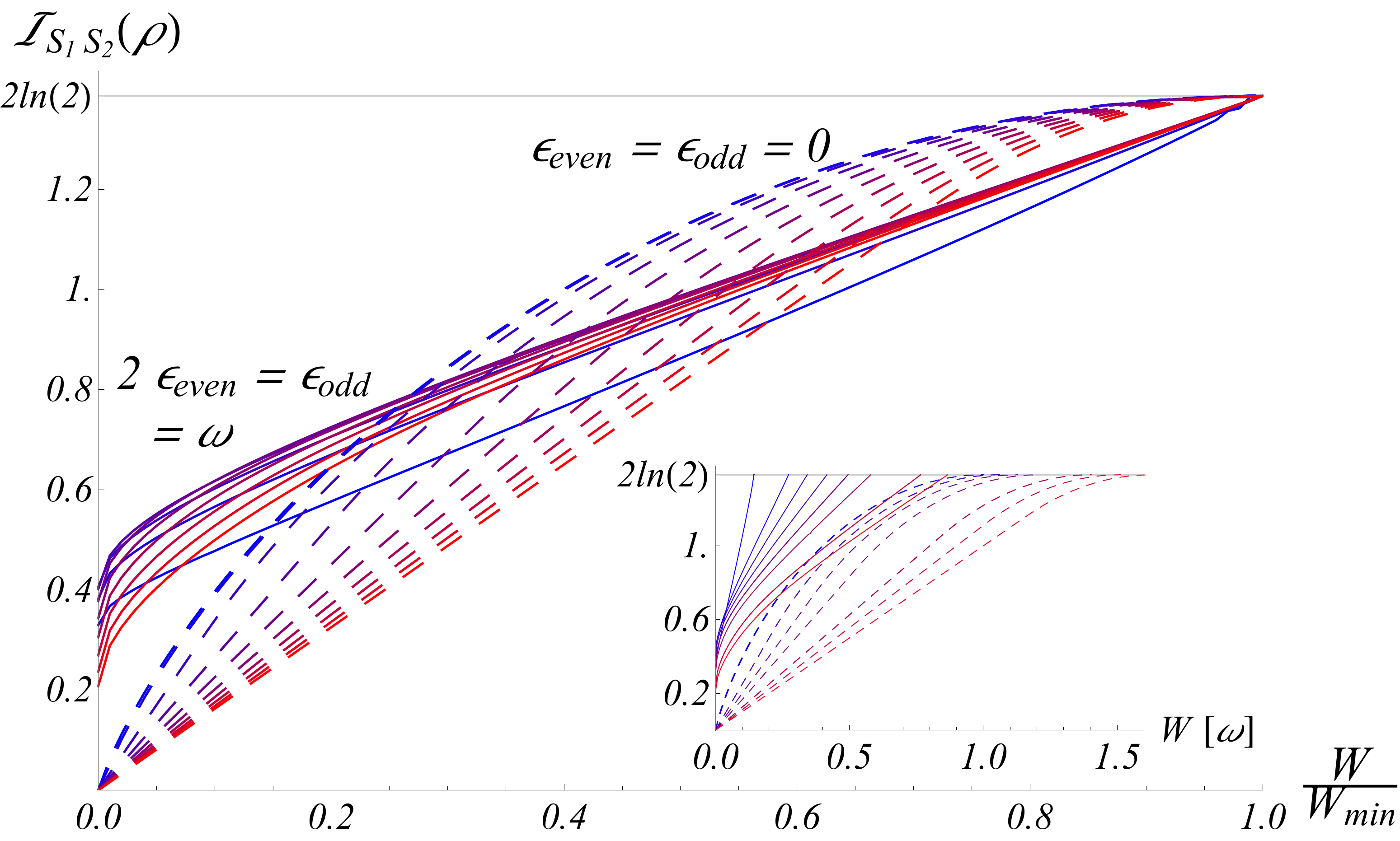}
(b)\includegraphics[width=0.445\textwidth,trim={0cm 2mm 0cm 0cm}]{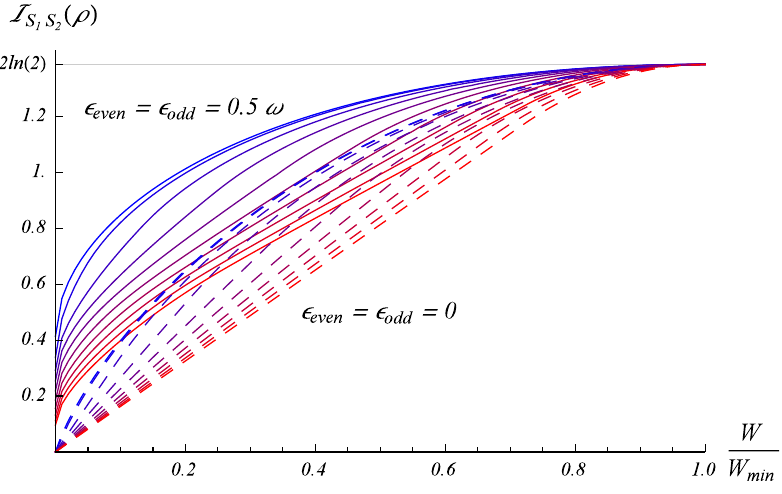}
(c)\includegraphics[width=0.445\textwidth,trim={0cm 2mm 0cm 0cm}]{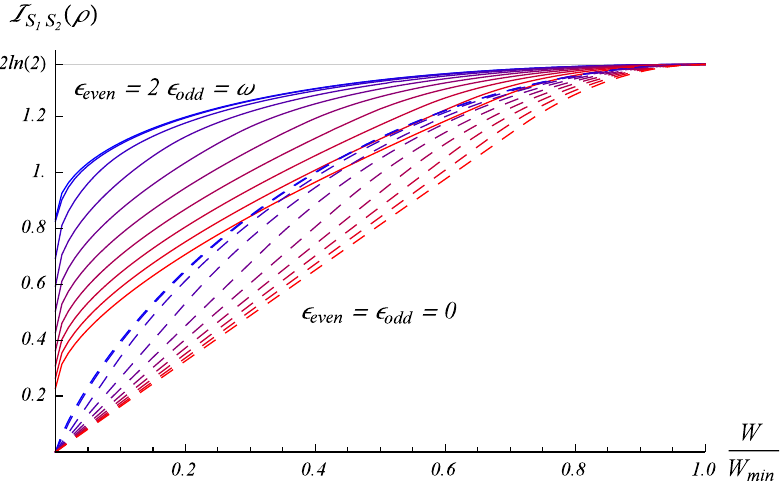}
\caption{
\textbf{Fermionic correlation cost:} The maximal correlation of the final state that is achievable for~a fixed input energy~$W$ [in units of $W_{\mathrm{min}}$ from Eq.~(\ref{eq:W min})] is shown for temperatures $T=0.1,\ldots,1$ (in units of $\hbar\omega/k_{B}$) in steps of $0.1$ (blue to red, top to bottom) for the ratios $\epsilon_{\mathrm{even}}/\epsilon_{\mathrm{odd}}=2,1$, and $0.5$ in (a),(b), and (c), respectively. In all cases, the achievable final correlation is larger in the presence of interactions (solid lines) than in their absence (dashed lines). In (a) this can be seen from the inset plot, where the horizontal axis is not scaled with $W_{\mathrm{min}}$. For (b) and (c) one may deduce this directly from the plots, since the solid lines are strictly above their corresponding dashed lines, and $W_{\mathrm{min}}$ is maximal when $\epsilon_{\mathrm{even}}=\epsilon_{\mathrm{odd}}=0$.}
\end{figure}

For a complete picture of the situation, it is also enlightening to study the behaviour of the total correlations $\mathcal{I}\Sonetwo(\rho)$ for interacting and noninteracting systems, as shown in Fig.~\ref{fig:fermionic generated correlation}. In all cases that we have considered, the presence of the interactions leads to~a larger amount of final state correlations $\mathcal{I}\Sonetwo(\rho)$, irrespective of the (relative) size and sign of the coupling constants.

\section{Two bosonic modes}\label{sec:two bosonic modes}

In this section we study the creation of correlations for two bosonic modes. We present an example for which we explicitly show that the overall final-state correlation $\mathcal{I}\Sonetwo(\rho)$ is always larger in the presence of interactions. This complements the observations made in Fig.~\ref{fig:fermionic generated correlation}, where we come to same conclusion using a numerical approach.

Let us now consider the simple, yet versatile system of two bosonic modes with creation and annihilation operators $\an{i}$ and $\andg{i}$ ($i=1,2$), respectively. The mode operators satisfy the canonical commutation relations $\comm{\an{i}}{\andg{j}}=\delta_{ij}$ and $\comm{\an{i}}{\an{j}}=0$. Such systems of two (or more) harmonics oscillators are of fundamental importance to quantum optics and quantum field theory. The correlations between bosonic modes have been extensively studied in continuous variable quantum information (see, e.g., Refs.~\cite{Weedbrooketal2012,AdessoRagyLee2014}) but they are also of interest in more specialized lines of research, such as, e.g., studies of entanglement in relativistic quantum field theory, see for example Refs.~\cite{FriisFuentes2013,BruschiLeeFuentes2013,BrownMartinMartinezMenicucciMann2013}. In addition to the usual free Hamiltonian
\begin{align}
    H\Sone\,+\,H\Stwo   &=\,\omega\,\bigl(\aonedg\aone\,+\,\atwodg\atwo\bigr)\,,
    \label{eq:bosonic modes free Hamiltonian}
\end{align}
where we have assumed that the two modes have the same frequency~$\omega$, we will consider the interaction term
\begin{align}
    \HI   &=\,\epsilon\,\bigl(\aone\atwo\,+\,\aonedg\atwodg\bigr)\,,
    \label{eq:bosonic modes interacting Hamiltonian}
\end{align}
with $\epsilon\in\mathbb{R}$. Since the system Hamiltonian $H\Sys=H\Sone+H\Stwo+\HI$ is quadratic in the mode operators, any thermal state~$\tau$ of $H\Sys$ is a Gaussian state that is fully described by its second moments, that is, its covariance matrix $\Gamma\Sys$ with components
\begin{align}
    \bigl(\Gamma\Sys\bigr)_{mn} &=\,\tr\bigl(\tau\bigl[\mathbb{X}_{m}\mathbb{X}_{n}+\mathbb{X}_{n}\mathbb{X}_{m}\bigr]\bigr)
    \,,
    \label{eq:covariance matrix components wrt to a modes}
\end{align}
with the quadrature operators $\mathbb{X}_{2n-1}=\bigl(\an{n}+\andg{n}\bigr)/\sqrt{2}$ and $\mathbb{X}_{2n}=-i\bigl(\an{n}-\andg{n}\bigr)/\sqrt{2}$. The first moments $\tr\bigl(\tau\mathbb{X}_{n}\bigr)$, which would normally also enter into Eq.~(\ref{eq:covariance matrix components wrt to a modes}), vanish for the state~$\tau$. This can easily be seen by diagonalizing $H\Sys$ using the Bogoliubov transformation
\begin{subequations}
\label{eq:bosonic bogo transform}
\begin{align}
    \cone   &=\,\cosh(u)\,\aone\,+\,\sinh(u)\,\atwodg\,,\\
    \ctwo   &=\,\cosh(u)\,\atwo\,+\,\sinh(u)\,\aonedg\,,
\end{align}
\end{subequations}
where $u=\tfrac{1}{2}\artanh(\epsilon/\omega)$, such that $\comm{\cn{i}}{\cndg{j}}=\delta_{ij}$ and $\comm{\cn{i}}{\cn{j}}=0$. With this transformation, the system Hamiltonian becomes
\begin{align}
    H\Sys   &=\,\tilde{\omega}\,\bigl(\conedg\cone\,+\,\ctwodg\ctwo\bigr)\,-\,2\hspace*{0.5pt}\tilde{\omega}\sinh^{2\hspace*{-0.5pt}}(u)\,,
    \label{eq:diagonalized bosonic Hamiltonian}
\end{align}
where $\tilde{\omega}=\sqrt{\omega^{2}-\epsilon^{2}}$. The eigenstates of $H\Sys$ are therefore the eigenstates of $\conedg\cone$ and $\ctwodg\ctwo$. Expanding the thermal state $\tau=\mathcal{Z}^{-1}e^{-\beta H\Sys}$ in terms of these eigenstates one quickly obtains $\tr\bigl(\tau\cn{i}\bigr)=\tr\bigl(\tau\cndg{i}\bigr)=0$. Since the Bogoliubov transformation relating the operators $\an{i},\andg{j}$ and $\cn{i},\cndg{j}$ is linear, this implies that also $\tr\bigl(\tau\an{i}\bigr)=\tr\bigl(\tau\andg{i}\bigr)=0$. The first moments vanish and $\Gamma\Sys$ completely describes the state $\tau$.

To assess the properties of the initial state, we then define the covariance matrix $\Gamma\Sys^{(c)}$ with respect to the operators $\cn{i}$, in complete analogy to Eq.~(\ref{eq:covariance matrix components wrt to a modes}). For the thermal state $\tau$, this $4\times4$ matrix is proportional to the identity, that is,
\begin{align}
    \Gamma\Sys^{(c)}    &=\,\nu(T)\mathds{1}_{4}\,=\,\coth(\tfrac{\tilde{\omega}}{2T})\,\mathds{1}_{4}\,,
\end{align}
where the identity $\mathds{1}_{4}$ is the covariance matrix of the pure two-mode vacuum state with respect to the operators $\cn{i}$. The mixedness of the state is hence captured solely by the prefactor $\nu(T)=\coth(\tfrac{\tilde{\omega}}{2T})$. The matrices $\Gamma\Sys^{(c)}$ and $\Gamma\Sys$ are related by a symplectic transformation $\mathcal{S}$ corresponding to the unitary Bogoliubov transformation of Eq.~(\ref{eq:bosonic bogo transform}), such that
\begin{align}
    \vspace*{-1mm}
    \Gamma\Sys  &=\,\mathcal{S}\,\Gamma\Sys^{(c)}\mathcal{S}^{T}\,.
\end{align}
The transformation $\mathcal{S}$ leaves the symplectic form $\Omega$, with components $\Omega_{kl}=-i\comm{\mathbb{X}_{k}}{\mathbb{X}_{l}}$ invariant, $\mathcal{S}\Omega\mathcal{S}^{T}=\Omega$. Consequently, also the eigenvalues $\nu(T)$ of $|i\Omega\Gamma\Sys|$, the symplectic eigenvalues
are left unchanged by the transformation $\mathcal{S}$. This means that, up to the prefactor $\nu(T)$, the covariance matrix $\Gamma\Sys=\nu(T)\mathcal{S}\mathcal{S}^{T}$ represents a pure two-mode state, which is hence locally equivalent to two-mode squeezed state~\cite{HolevoWerner2001}. Due to the presence of $\nu(T)$, the overall state is nonetheless mixed and correlated, but may or may not be entangled, depending on the size of $\nu(T)$~\cite{BruschiFriisFuentesWeinfurtner2013,BruschiPerarnauLlobetFriisHovhannisyanHuber2015}.

Any available energy~$W$ may then be used to further correlate the system by a combination of cooling [i.e., reducing $\nu(T)$] and two-mode squeezing along the direction in phase space corresponding to the two-mode squeezed state $\mathcal{S}\mathcal{S}^{T}$. These transformations leave the subsystems in local thermal states with respect to $H\Sone$ and $H\Stwo$ and therefore optimally correlate\footnote{Note that two-mode squeezing is generally not the optimal entangling transformation and may be outperformed by non-Gaussian transformations~\cite{BruschiPerarnauLlobetFriisHovhannisyanHuber2015}.} the subsystems at any given work cost~\cite{HuberPerarnauHovhannisyanSkrzypczykKloecklBrunnerAcin2015}. Consequently, the presence of the interaction Hamiltonian~$\HI$ is here equivalent to an increased energy supply in the noninteracting case, and the overall correlations $\mathcal{I}\Sonetwo(\rho)$ are always larger than in the noninteracting case at a fixed work cost~$W$.

The correlations that can in principle be generated in this infinite-dimensional Hilbert space are unbounded. However, the energy cost of creating additional correlations increases as the state becomes more correlated. The newly generated correlations~$\Delta\mathcal{I}\Sonetwo$ may hence be more or less expensive than in the noninteracting case, depending on the initial temperature, coupling strength~$\epsilon$, and the available energy.

\section{Conclusion}\label{sec:discussion}

We have investigated the work cost of creating correlations between interacting quantum systems and compared our results to previous studies~\cite{HuberPerarnauHovhannisyanSkrzypczykKloecklBrunnerAcin2015,BruschiPerarnauLlobetFriisHovhannisyanHuber2015} of the correlation cost in noninteracting systems. While the notion of isolated, noninteracting systems may appear more natural from the perspective of quantum communication scenarios, our approach here is motivated by the ubiquity of interactions present in nature. Hence, assuming that the presence of the interactions cannot be avoided or controlled, we find that the interactions can nonetheless be harnessed.

For such naturally occurring interactions we have identified general strategies for finite-dimensional systems to reduce the energy cost of creating correlations. These strategies, which apply to any finite-dimensional bipartite system with arbitrary interaction Hamiltonian, improve on previous bounds for non-interacting systems, at least in some low-energy regime. Nevertheless, the exact relation between the interactions and the correlation cost is complicated. The work cost of correlations strongly depends on the exact configuration of the interaction terms and thus on the underlying physics. To illustrate the general strategies, we therefore choose some exemplary physical systems \textemdash\ qubits, as well as fermionic and bosonic modes \textemdash\ to showcase the usefulness of the interactions.

In our examination, we have focused on the mutual information as a measure of the generated correlations, capturing both classical and genuine quantum correlations. The notoriously difficult case of characterizing the cost of entangling interacting quantum systems, which would provide further insight into the relation between the practically motivated resource theories of QI and QT, is hence left open for future investigation.

Moreover, while we have here considered arbitrary operations on the system, inevitable noise and practical design may favor operations that can be directly implemented through the natural interactions present in the underlying systems. It would hence be interesting to compare such physically motivated protocols with the optimal protocols derived here, including also noncyclic processes where the interactions can be switched on or off at will. Further open questions include the general cost and impact of interactions on single-shot information processing capabilities.

\begin{acknowledgments}
We thank Hans J. Briegel and Vedran Dunjko for fruitful comments. N.~F. is grateful to Universitat Aut\`{o}noma de Barcelona and the LIQID collaboration for hospitality and acknowledges funding by the Austrian Science Fund (FWF) through the SFB FoQuS: F4012. M.~H. acknowledges funding from the Swiss National Science Foundation (AMBIZIONE PZ00P2\_161351), from the Spanish MINECO through Project No.~FIS2013-40627-P and the Juan de la Cierva fellowship (JCI 2012-14155), from the Generalitat de Catalunya CIRIT Project No.~2014 SGR 966, and from the EU STREP-Project “RAQUEL". M.~P.-L. acknowledges funding from the Spanish MINECO (FOQUS FIS2013-46768-P and SEV-2015-0522), the grant No.~FPU13/05988, and the Generalitat de Catalunya (SGR 875). M.~P.-L. and M.~H. are grateful for support from the EU COST Action No.~MP1209, "Thermodynamics in the quantum regime".
\end{acknowledgments}

\newpage
\appendix*
\section{Correlations in a fermionic thermal state}\label{sec:CorrelationsFermions}

\begin{figure}[ht!]
\label{fig:fermionic thermal state correlation}
(a)\includegraphics[width=0.45\textwidth,trim={0cm 0mm 0cm 0mm}]{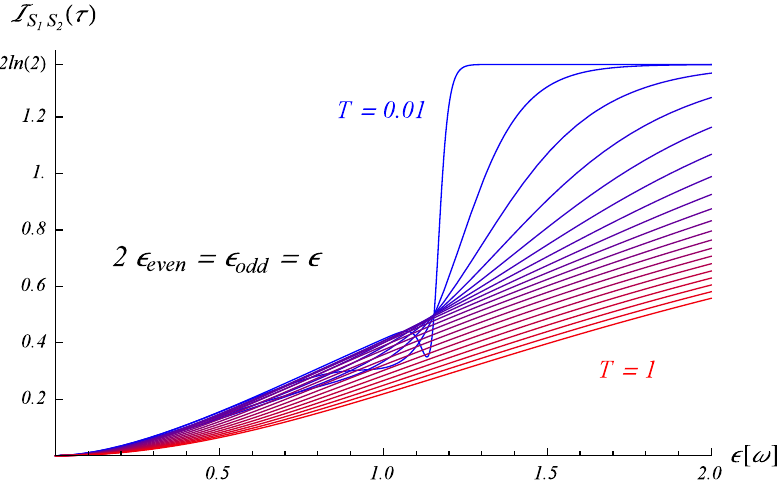}
(b)\includegraphics[width=0.45\textwidth,trim={0cm 1mm 0cm 0mm}]{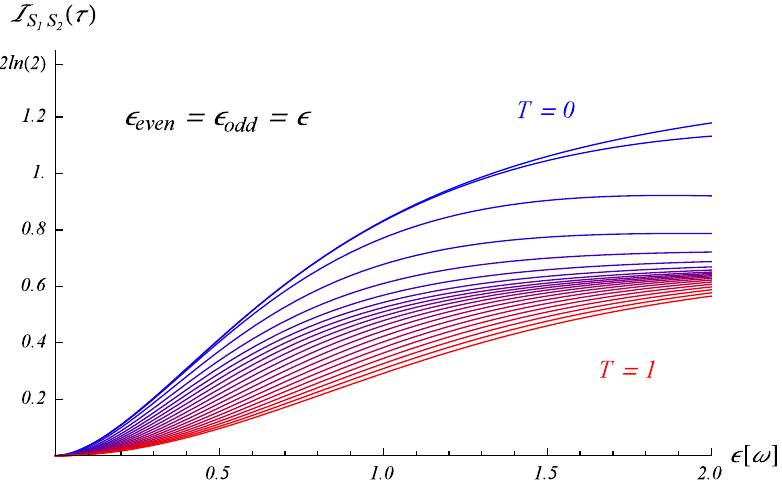}
(c)\includegraphics[width=0.45\textwidth,trim={0cm 2mm 0cm 0mm}]{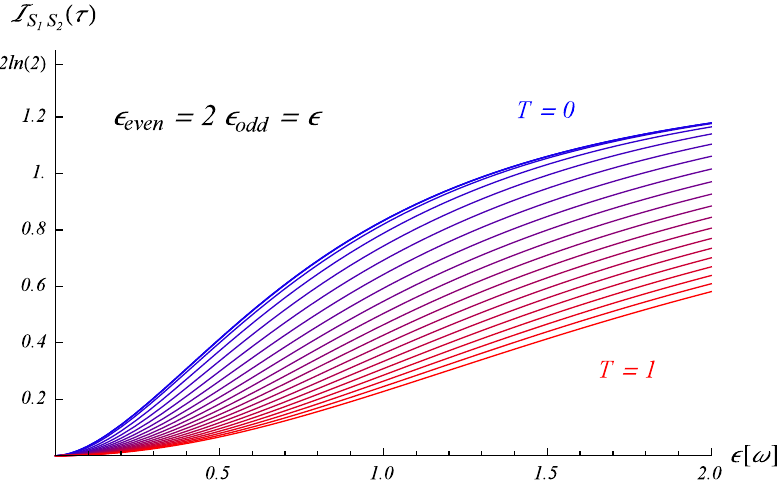}
\caption{
\textbf{Fermionic thermal state correlation:} The correlation of the initial thermal state $\tau(\beta)$, as measured by the mutual information $\mathcal{I}\Sonetwo$, depend on the (relative) and absolute sizes of the couplings $\epsilon_{\mathrm{even}}$ and $\epsilon_{\mathrm{odd}}$. The correlation is plotted for temperatures $T=0(0.01),\ldots,1$ (in units of $\hbar\omega/k_{B}$) in steps of $0.05$ (blue to red, top to bottom) for the ratios $\epsilon_{\mathrm{even}}/\epsilon_{\mathrm{odd}}=2,1$, and $0.5$ in (a),(b), and (c), respectively.}
\end{figure}

Here we study the correlations present in the initial thermal state $\tau_{\beta}$ of two fermionic modes, given by Eq.~(\ref{eq:fermionic thermal state}). Whereas the thermal states of the noninteracting Hamiltonian (i.e., for $\epsilon_{\mathrm{even,odd}}=0$) are uncorrelated, the thermal states of interacting Hamiltonians feature some correlations. For instance, consider the ground state, that is, the limit $T\rightarrow0$ $(\beta\rightarrow\infty)$, which arises as the eigenstate of the Hamiltonian with the smallest eigenvalue. As can be seen from Eq.~(\ref{eq:fermion eigenvalues}), this can be either $\fket{\lambda_{3}}$ or $\fket{\lambda_{4}}$, depending on the relative sizes of $|\epsilon_{\mathrm{odd}}|$ and $\sqrt{\omega^{2}+\epsilon_{\mathrm{even}}^{2}}$. That is,
\begin{align}
 \tau(\beta\rightarrow\infty)   &=\,
    \begin{cases}
        \rho_{3} &   \mbox{if}\ \ |\epsilon_{\mathrm{odd}}|\,>\,\sqrt{\omega^{2}+\epsilon_{\mathrm{even}}^{2}}\\
        \frac{1}{2}\bigl(\rho_{3}\,+\,\rho_{4}\bigr) &   \mbox{if}\ \ |\epsilon_{\mathrm{odd}}|\,=\,\sqrt{\omega^{2}+\epsilon_{\mathrm{even}}^{2}}\\
        \rho_{4} &   \mbox{if}\ \ |\epsilon_{\mathrm{odd}}|\,<\,\sqrt{\omega^{2}+\epsilon_{\mathrm{even}}^{2}}
    \end{cases}\,,
\end{align}
with $\rho_{3}=\fket{\lambda_{3}}\!\fbra{\lambda_{3}}$ and $\rho_{4}=\fket{\lambda_{4}}\!\fbra{\lambda_{4}}$. Both $\fket{\lambda_{3}}$ and $\fket{\lambda_{4}}$ are correlated, but only the former state is maximally correlated. It is hence expected that the relative sizes of the coupling constants strongly influence the initial amount of correlations, see Fig.~\ref{fig:fermionic thermal state correlation}. To evaluate the mutual information of Eq.~(\ref{eq:mutual inf definition}), we still need to specify the reduced density operators. These are found to be diagonal, with matching matrix elements, i.e.,
\begin{subequations}
\label{eq:fermions thermal state marginals}
\begin{align}
    \tau\Sone(\beta) &=\,\tfrac{1}{2}(1+\tau_{0})\,\fket{0}\!\fbra{0}\,+\,\tfrac{1}{2}(1-\tau_{0})\,\fket{1_{1}}\!\fbra{1_{1}}\,,
    \label{eq:fermions thermal state marginals 1}\\
    \tau\Stwo(\beta) &=\,\tfrac{1}{2}(1+\tau_{0})\,\fket{0}\!\fbra{0}\,+\,\tfrac{1}{2}(1-\tau_{0})\,\fket{1_{2}}\!\fbra{1_{2}}\,,
    \label{eq:fermions thermal state marginals 2}
\end{align}
\end{subequations}
with the coefficient $\tau_{0}$ given by
\begin{align}
    \tau_{0}    &=\,\frac{\omega\,\sinh\bigl(\beta\sqrt{\omega^{2}+\epsilon_{\mathrm{even}}^{2}}\,\bigr)}{\sqrt{\omega^{2}+\epsilon_{\mathrm{even}}^{2}}
    \Bigl(\cosh\bigl(\beta\epsilon_{\mathrm{odd}}\bigr)+\cosh\bigl(\beta\sqrt{\omega^{2}+\epsilon_{\mathrm{even}}^{2}}\bigr)\Bigr)}\,.
    \label{eq:fermions thermal marginals coefficients}
\end{align}
With the eigenvalues of the thermal state given by $e^{-\beta\lambda_{i}}$ and those of the marginals by $\tfrac{1}{2}(1\pm\tau_{0})$ one can then easily evaluate the entropies~$S(\tau)$, $S(\tau\Sone)$ and $S(\tau\Stwo)$, and hence the mutual information, shown in Fig.~\ref{fig:fermionic thermal state correlation}.

\newpage

\end{document}